# Studying the Scientific Mobility and International Collaboration Funded by the China Scholarship Council

Zhichao Fang[1], Wout Lamers[1], and Rodrigo Costas[1,2]

[1] *z.fang@cwts.leidenuniv.nl, w.s.lamers@cwts.leidenuniv.nl, rcostas@cwts.leidenuniv.nl*
Centre for Science and Technology Studies (CWTS), Leiden University, Wassenaarseweg 62A, Leiden, 2333AL
(The Netherlands)

[2] DST-NRF Centre of Excellence in Scientometrics and Science, Technology and Innovation Policy,
Stellenbosch University (South Africa)

**Abstract**
Every year many scholars are funded by the China Scholarship Council (CSC). The CSC is a funding agency established by the Chinese government with the main initiative of training Chinese scholars to conduct research abroad and to promote international collaboration. In this study, we identified these CSC-funded scholars sponsored by the China Scholarship Council based on the acknowledgments text indexed by the Web of Science. Bibliometric data of their publications were collected to track their scientific mobility in different fields, and to evaluate the performance of the CSC scholarship in promoting international collaboration by sponsoring the mobility of scholars. Papers funded by the China Scholarship Council are mainly from the fields of natural sciences and engineering sciences. There are few CSC-funded papers in the field of social sciences and humanities. CSC-funded scholars from mainland China have the United States, Australia, Canada, and some European countries, such as Germany, the UK, and the Netherlands, as their preferential mobility destinations across all fields of science. CSC-funded scholars published most of their papers with international collaboration during the mobility period, with a decrease in the share of international collaboration after the support of the scholarship.

# Introduction

The China Scholarship Council (CSC), a non-profit funding organization entrusted by the Ministry of Education of the People's Republic of China, was established in 1996[1] for the award, enrolment and administration of a series of Chinese Government Scholarship programs. These scholarship programs were set up by the Chinese government to encourage and sponsor Chinese students and scholars to study and conduct research abroad, as well as to sponsor international students and scholars to study and do research in Chinese universities[2]. According to the *provisions* issued by the Ministry of Education of China in 2007[3], the China Scholarship Council is in charge of selecting and dispatching students and scholars based on their own applications and organized experts review, and the CSC scholarship awardees are obligated to move to their targeted visiting countries and stay there to study or conduct research within the prescribed time. Moreover, for sponsored Chinese students and scholars, they are required to return to China and work for at least two years following completion of sponsorship by principle. Therefore, it is a kind of circular transnational scientific mobility (Jöns, 2007) supported by the government with specific initiatives: to train talents and promote international collaboration in some important fields, and lastly, to implement the so-called *return brain drain* (Jonkers & Tijssen, 2008). As Cao (2008) suggested, although a small but growing return migration of Chinese researchers has been seen, the whole return rate is low and many highly qualified academics still stay abroad for multiple reasons.

Therefore, the CSC scholarship is expected to play an important role in training scholars with international research experience and attracting them back to China.

Scientific mobility has been related to higher scientific impact of papers and scientists themselves (Wagner & Jonkers, 2017; Sugimoto et al., 2017; Halevi, Moed, & Bar-Ilan, 2015). Cañibano (2017) conceived it as a mechanism for the allocation of human resources in research labour market, through which *brain gain*, *brain drain*, and *brain circulation* are happening and global inequality is therefore increased (Scott, 2015). However, different from unprompted brain drain, as Cañibano & Woolley (2015) pointed out, there are numerous developing countries formulated grant policies with *train* and *attract back* rationales. The China Scholarship Council is a typical example of national funding organisations implementing this train and attack back policy, together with some countries in Latin America, such as Peru[4], which are suffering increasing brain drain too (Adams Jr, 2003). Those scholarship awardees were expected to gain international experience (Ackers, 2008) in leading countries in their subject fields, and expand transnational collaboration networks through mobility (Meyer, Kaplan, & Charum, 2001; Cañibano, Fox, & Otamendi, 2015). Jonkers and Tijssen (2008) found that the overseas experience of Chinese plant molecular life researchers who returned to their home country does have a distinct positive impact on the publication productivity; moreover, it was observed a positive correlation between researchers' overseas experience and the quantity of their corresponding transnational co-publications. Based on survey data, Scellato, Franzoni, & Stephan (2015) concluded that migrants and returnees hold larger international research networks compared to those native researchers lacking an international background. The positive effect of international mobility among countries with rich research environments on qualified international collaboration was also observed by Kato & Ando (2016).

Identifying the mobility of scientists is the first step to quantitatively analyse and evaluate the behaviours and the following impact of mobile scientists. Bibliometric data, especially the affiliation data of authors, have been widely used in identifying and tracing the trajectory of individual authors since Laudel (2003). Through this bibliometric approach, the relationship between researcher mobility and other bibliometric indicators, such as their collaborative networks, paper productivity and citations, were discussed in previous studies (Furukawa, Shirakawa, & Okuwada, 2011; Aksnes et al., 2013). In addition, on the basis of the development and application of large-scale author name disambiguation algorithm (Caron & van Eck, 2014), studies on scientific mobility became more extensive and worldwide. For instance, Scopus author-affiliation linking and author profiling were used by Moed, Aisati & Plume (2013) and Moed & Halevi (2014) to track international migration. Based on the Web of Science data, Robinson-Garcia et al. (2019) presented a taxonomy consisting of four mobility types of scientists: not mobile researchers, directional travellers, non-directional travellers, and migrants, by using instances of changes in (or multiplicity of) affiliations for a single scholar as the proxy for identifying mobility. The same method was applied by Sugimoto et al. (2017) and Chinchilla-Rodríguez et al. (2018) in tracking international mobility, under the background of controversial travel bans issued by the US government which led to a negative influence on the scientific activities of scholars from those restricted countries (Reardon, 2017; Morello & Reardon, 2017). However, there are some limitations to this methodology based on the changes of authors' affiliations, such as ignoring the scientific mobility without research outputs and underrepresenting short-term stays without changes of authors' affiliations (Robinson-Garcia et al., 2019).

The mobility and effect resulted by some international scholarship programs have been analysed based on annual report or survey data, such as the US Fulbright Program (Kahn & MacGarvie, 2011). In this paper, we study the mobility of scholars who were funded by the China Scholarship Council by mining the acknowledgements text of the Web of Science papers. Their mobility destinations were identified through affiliation information of authorships, and all of their Web of Science publications were collected by using the author name disambiguation algorithm by Caron & van Eck (2014) to investigate the proportion of their papers with international collaboration in different periods (before, during, and after sponsorship of the CSC scholarship). The scientific mobility of Chinese CSC-funded scholars is assumed to reflect the initiatives of government for training elite academics with international experience and establishing closer international collaboration. The performance of this funding policy has significant implications for the funder agency and policy makers. The main objective of this study is to investigate the situation and performance of the China Scholarship Council in promoting brain exchange and international collaboration. Here we addressed the following research questions:

- Firstly, how is the distribution of CSC-funded papers in different subject fields? Which fields have the most papers funded by the China Scholarship Council?
- Secondly, based on the affiliation information of their research outputs, where did identified CSC-funded scholars choose to get training and establish collaboration in different fields?
- Lastly, to what extent the CSC-funded scholars engaged in international collaboration before, during, and after being supported by the China Scholarship Council?

**Data and methods**

*Papers funded by the China Scholarship Council*

According to the *Agreement on Funding the Study Abroad* that scholars have to sign with the China Scholarship Council when they are awarded the scholarship supporting them to visit or study abroad[5], they are required to acknowledge the funding from the China Scholarship Council in their research outputs conducted during the period of sponsorship. Therefore, on the basis of the published funding acknowledgements which have been indexed by the Web of Science (WoS) since 2008 for science and medicine papers, and since 2015 for social science papers (Paul-Hus, Desrochers, Costas, 2016), respectively, WoS papers funded by the China Scholarship Council can be identified.

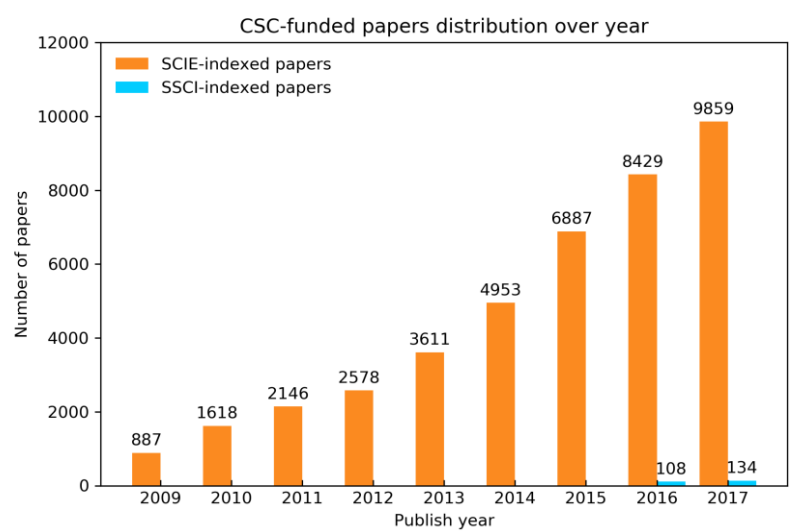

**Figure 1. Temporal distribution of the CSC-funded papers.**

By using the in-house version of the Web of Science maintained at CWTS of Leiden University, we collected 40,968 SCIE-indexed papers published from 2009 to 2017 and 242 SSCI-indexed papers published from 2016 to 2017 with the China Scholarship Council or its variations (such as 'the CSC Scholarship' and 'the Chinese Scholarship Council') listed in their funding sources and standardized in the 'CWTS Thesaurus' (Van Honk, Calero-Medina, & Costas, 2016). Only the document type of Article and Review are considered. The temporal distribution of these 41,210 papers is shown in Figure 1. From 2009 onwards, the number of papers funded by the China Scholarship Council increased over time. In the period from 2016 to 2017, when both SCIE-indexed papers and SSCI-indexed papers have recorded funding sources, the quantity of SCIE-indexed papers acknowledging the China Scholarship Council is much higher than SSCI-indexed papers.

*Author disambiguation and identification*

In this study 41,210 CSC-funded papers are contributed by 228,365 authors in total. In order to study the mobility and performance of scholars who were actually sponsored by the China Scholarship Council, firstly it is necessary to disambiguate author names to solve the problems caused by homonymy and name variants (Costas, van Leeuwen, & Bordons, 2010). A large-scale author name disambiguation algorithm developed by Caron and van Eck (2014), which has been implemented in the in-house CWTS version of the Web of Science, was used to identify unique scholars authoring scientific papers. Based on the disambiguation algorithm, we found that 41,210 CSC-funded papers were contributed by 141,614 unique individuals. However, not all of them were funded by the China Scholarship Council and supported by the Chinese government to move. Actually, most contributed individuals are co-authors of CSC-funded scholars. Therefore, in order to explore the scientific mobility and international collaboration of CSC-funded scholars, we identified those real CSC Scholarship awardees among all authors through mining the acknowledgements text, by matching the authors' names in the acknowledgements. A brief description of this text mining methodology is as follows:

- Firstly, specific full sentences containing "China Scholarship Council" or its variants, such as "CSC scholarship", "Chinese Scholarship Council", are extracted from the acknowledgement text of each CSC-funded paper. Thus, an example of an extracted full sentence is "Long Chen is supported by a scholarship from the China Scholarship Council (CSC)." (extracted from the acknowledgements of WoS paper: 000373106500002).
- Secondly, for every CSC-funded paper, various forms of its authors' names are developed (e.g. full name and last name, initials and surnames, etc.). Every author is also assigned with an ordinal number based on their sequences, such as "the first author", "the second author". Taking the WoS paper (000373106500002) as an example, there are two authors: "Bernreuther, Werner" and "Chen, Long". Possible forms of each author's name in the acknowledgements are developed as: Bernreuther, W ("Bernreuther Werner", "Werner Bernreuther", "Bernreuther. W", "B.W.", "BW", "the first author", etc.), Chen, L ("Chen Long", "Long Chen", "Chen. L", "C.L.", "CL", "the second author", etc.).
- Lastly, for each paper, the developed authors' names, together with their ordinal numbers, are matched with the corresponding extracted full funding sentence to find out if they appear in it or not. Once an author's name or ordinal number is matched, that author would be identified as the CSC-funded scholar of that paper. In the above example, the name of "Long Chen" was matched in the full funding sentence, so Chen, L was identified as the CSC-funded scholar.

Finally, among 141,614 unique authors, 9,562 of them were identified as CSC-funded scholars, contributing to 16,037 unique papers (i.e. for around 39% of the CSC-funded publications we identified the funded scholar).

*Papers contributed by CSC-funded scholars*

For the 9,562 identified CSC-funded scholars, all of their published WoS papers until March, 2018 were collected based on the disambiguation algorithm and further cleaned by authors' first names, affiliations, and co-authorship. In addition to the 16,037 CSC-funded papers with identified scholars, there are other 69,708 WoS papers that were authored by these scholars, thus totalling 85,745 (i.e. 16,037 CSC-funded publications with identified scholars and their other 69,708 publications that were not sponsored by the CSC scholarship or did not mention their names in the acknowledgements text).

By using the created dates of DOIs collected from Crossref as the proxy for the precise publication dates of papers (when the created date was not available (account for 9.6%), the WoS publication year was used as the alternative). As a result the 85,745 papers were classified into three periods: *before sponsorship*, *during sponsorship*, and *after sponsorship*:

- For papers with funding acknowledgements containing the China Scholarship Council, they are classified as *during sponsorship* (19,328 papers, account for 22.5%). Among these 19,328 papers, 16,037 of them mentioned the specific identified authors' names in the acknowledgements texts, while others only listed the CSC as a funding source without mentioning the CSC-funded scholars' names;
- For papers without funding acknowledgements containing the China Scholarship Council, the output of the identified CSC-funded scholars is analysed in order to estimate the first and last publications of the scholar with a CSC funding acknowledgment. Thus, those papers with publication date earlier than the first CSC-funded papers are classified as *before sponsorship* (30,399 papers, account for 35.5%), while papers with publication date later than the last CSC-funded papers are classified as *after sponsorship* (23,935 papers, account for 27.9%).

If a paper was authored by more than one identified CSC-funded scholars, as long as its acknowledgements contain the China Scholarship Council, it would be *during sponsorship*. Otherwise, it would be identified as *before sponsorship* or *after sponsorship* only if it could be classified into a specific period in all cases of scholars. There are two types of not CSC-funded papers that were excluded from this analysis (12,083 papers in total, accounting for 14.1%) since their classification is ambiguous based on the publication date. One is a group of papers that cannot unambiguously classified into before or after sponsorship since they were contributed by more than one CSC-funded scholar with different sponsorship periods, and they can be classified into different periods based on different authors' sponsorship periods. For instance, if a paper was authored by two identified CSC-funded scholars, from the view of the first author, it should be classified as *before sponsorship*, but from the view of the second author, it is *after sponsorship*, then this paper was excluded; the other group of excluded papers is that of papers published between the first and last publication dates of identified CSC-funded papers, namely the publication date of a paper is during sponsorship period but it didn't acknowledge the China Scholarship Council. There are some possible reasons for this situation. For example, if a paper was completed before sponsorship but published during sponsorship due to the publication delay, its acknowledgements would not contain the CSC scholarship. Besides, if the main work of a paper was not conducted during sponsorship, it is not necessary to acknowledge the China Scholarship Council even though it was published during sponsorship.

## Results

*Field distribution of CSC-funded papers*

According to the outline[6] for selecting and dispatching CSC-funded scholars launched by the China Scholarship Council, candidates from the key fields that highlighted by two Chinese government policy documents (listed in Table 1) and humanities and social sciences have the priority to be funded. Key fields of natural sciences and engineering sciences account for the majority in these two Chinese government national outlines, especially those fields play significant roles in industry development, for instance, Manufacturing, Energy, Materials, and Transportation. Biotechnology, Environment, and Agriculture, which are of great concern to population health and public security, are also highlighted by these two policy documents. Some social science related fields are underlined as key fields too, such as Financial Accountancy, Education, and Politics and Law.

**Table 1. Key fields stated in two Chinese Government National Outlines.**

| | *National Outline for Medium and Long-Term Talents Development Plan (2010-2020)* | *National Outline for Medium and Long-Term Science and Technology Development Plan (2006-2020)* |
|---|---|---|
| Key fields | Equipment Manufacturing<br>Information<br>Biotechnology<br>Advanced Materials<br>Aviation and Astronautics<br>Ocean<br>Financial Accountancy<br>International Business<br>Ecology and Environment Protection<br>Energy Resources<br>Modern transportation<br>Agricultural Science and Technology<br>Education<br>Politics and Law<br>Propaganda, Ideology and Culture<br>Disaster Prevention and Reduction | Energy<br>Water and Mineral Resources<br>Environment<br>Agriculture<br>Manufacturing Industry<br>Transportation Industry<br>Information and Modern Service Industry<br>Population and Health Sciences<br>Urbanization and Urban Development<br>Public Security<br>National Defence |

Figure 2 shows the field distribution of 41,210 CSC-funded papers and 16,037 of them with identified CSC-funded authors. Fractional-counting was applied to calculate the distribution of papers across fields when the paper belongs to more than one subject field. The classification of fields and field weights of each paper are based on the NOWT classification system (Tijssen, Hollanders, & van Steen, 2010) developed by CWTS. For most subject fields, nearly 40% of CSC-funded papers have authors identified as CSC-funded scholars according to the detailed statement in the funding acknowledgements. Physics and Materials Science is the field with most CSC-funded papers, followed by Chemistry and Chemical Engineering. Nearly 40% of total CSC-funded papers belong to these two fields, and the quantity is around twice larger than the third most productive field: Basic Life Sciences. Most key fields presented in Table 1 have a considerable number of CSC-funded research outputs, except for social sciences fields, although some social sciences are regarded as the key fields by the outline.

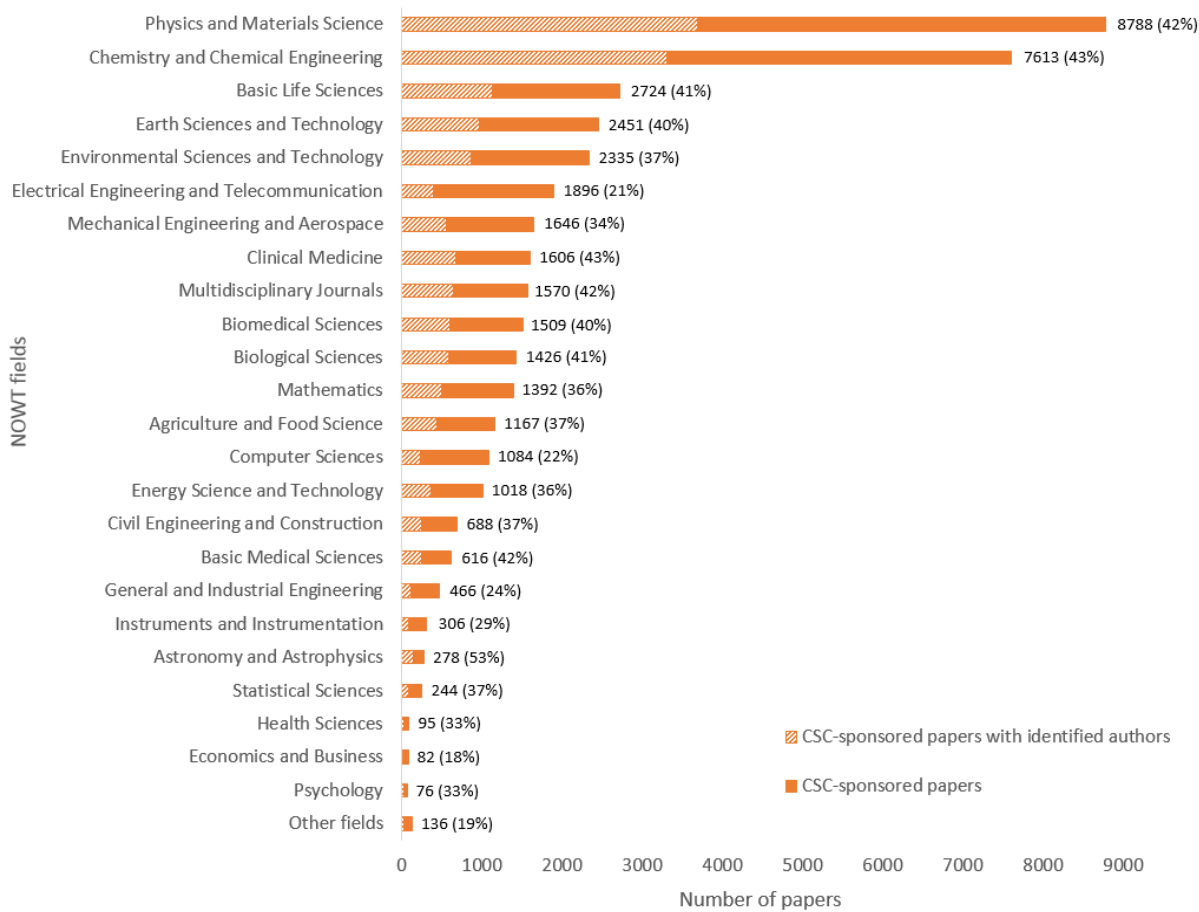

**Figure 2. Field distribution of CSC-funded papers.**

*Scientific mobility of Chinese CSC-funded scholars*

In this section we study the countries or regions where CSC-funded scholars applied and admitted to receive academic training or conduct research. Their mobility destinations not only reflect the sponsored scholars' intentions to move, but also reveal the recognition from expert reviewers organized by the China Scholarship Council to the research quality of that country in specific fields. In this study we considered both identified CSC-funded authors' affiliations and their co-authors' affiliations to identify their mobility destinations. Only identified authors from mainland China were taken into consideration to investigate their mobility destinations under the support of the China Scholarship Council, because for CSC-funded authors from other countries, their mobility destination must be China in this case. We identified Chinese CSC-funded authors by matching if they have a typical Chinese last name from mainland China (9,467 Chinese individuals were extracted, accounting for 99% of all identified CSC-funded scholars). The methodology for identifying mobility destinations are following:

- If the CSC-funded author has only one affiliated country and it is not China, then this country is the mobility destination (3,052 authors, account for 32.2%);
- If the CSC-funded author has two affiliated countries and one of them is China, then the other one is the mobility destination (4,515 authors, account for 47.7%);
- If the CSC-funded author is only affiliated to China or the author does not have affiliation information, but the co-authors are affiliated to just one another country, than that country is the mobility destination (1,233 authors, account for 12.9%);

However, there are some cases that the mobility destination cannot be identified:

- If the CSC-funded author was affiliated to more than one country except China (147 authors, account for 1.6%);
- If the CSC-funded author is only affiliated to China, while the co-authors are only affiliated to China too or affiliated to more than one country except China (814 authors, account for 8.6%).

For 8,800 Chinese CSC-funded scholars with publications whose mobility destinations can be identified more accurately as depicted above, their mobility routes are presented in Figure 3. The United Sates is the main mobility destination, together with some developed European countries, such as Germany, the UK, France, and the Netherlands. Scholars also prefer to move to Canada, Australia, and Japan to conduct research and obtain international experience.

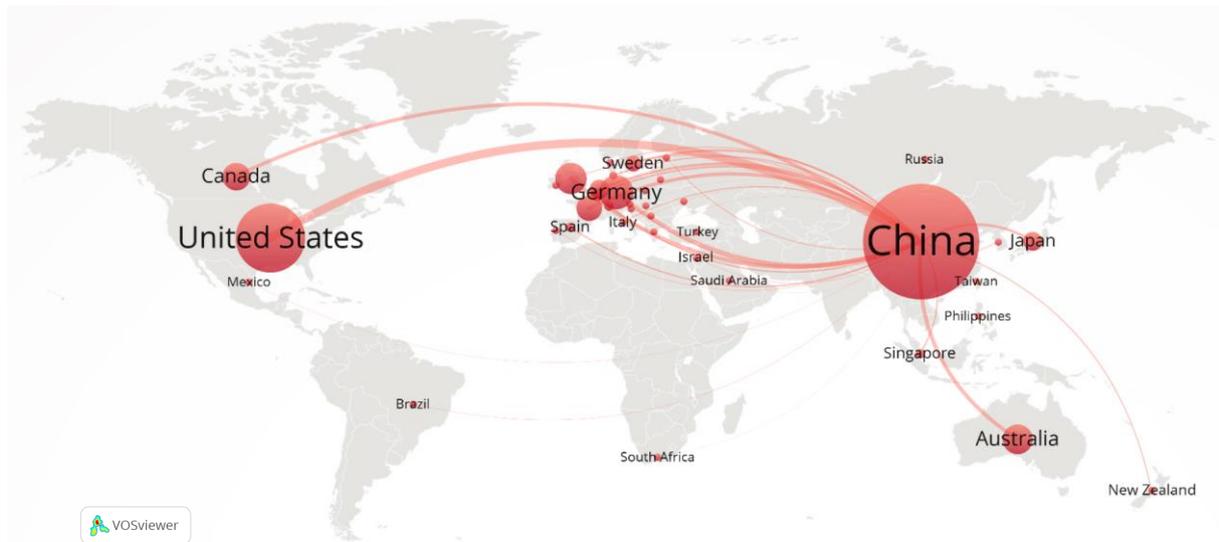

**Figure 3. Scientific mobility destinations of CSC-funded scholars from mainland China.**

**Table 2. Top 10 mobility destinations across five subject fields.**

| Rank | Physical Sciences and Engineering | Biomedical and Health Sciences | Life and Earth Sciences | Mathematics and Computer Sciences | Social Sciences and Humanities |
|---|---|---|---|---|---|
| 1 | US | US | US | US | US |
| 2 | Germany | Germany | Germany | UK | Netherlands |
| 3 | UK | Canada | Australia | Canada | UK |
| 4 | Australia | UK | Canada | Australia | Australia |
| 5 | France | Australia | UK | France | Germany |
| 6 | Canada | Netherlands | Netherlands | Germany | Canada |
| 7 | Japan | France | France | Netherlands | Spain |
| 8 | Sweden | Sweden | Japan | Japan | France |
| 9 | Netherlands | Japan | Belgium | Singapore | Denmark |
| 10 | Belgium | Belgium | Sweden | Spain | Singapore |

The top 10 mobility destinations across five main disciplines (based on the CWTS classification system developed by Waltman & Van Eck, (2012)) are listed in Table 2. The field that a scholar belongs to was based on the publications during the sponsorship. Fractional counting was employed when their publications clustered into various fields. In all fields, the US is always the most preferential destination for CSC-funded scholars with research outputs. Followed by Germany, another main choice in the fields of natural sciences and engineering sciences. Canada, Australia, and other European countries, such as the UK, the Netherlands, France, Sweden, and Belgium, also serve as important destinations of scientific mobility funded by the China Scholarship Council. Japan is the most preferential Asian country in most fields. Singapore, another Asian country, ranks in the top 10 in the field of Mathematics and Computer Sciences, and Social Sciences and Humanities.

*The performance of the CSC scholarship in promoting international collaboration*

Promoting international collaboration is one of the most important goals of the China Scholarship Council for sponsoring scientific mobility. We calculated the indicator *PP(IC)*, namely the proportion of papers with international collaboration (papers with affiliations from more than one country), to measure the transnational collaboration situation of papers contributed by identified CSC-funded scholars before, during, and after the sponsorship period. Figure 4 presents the proportion of papers with international collaboration across fields in different sponsorship periods.

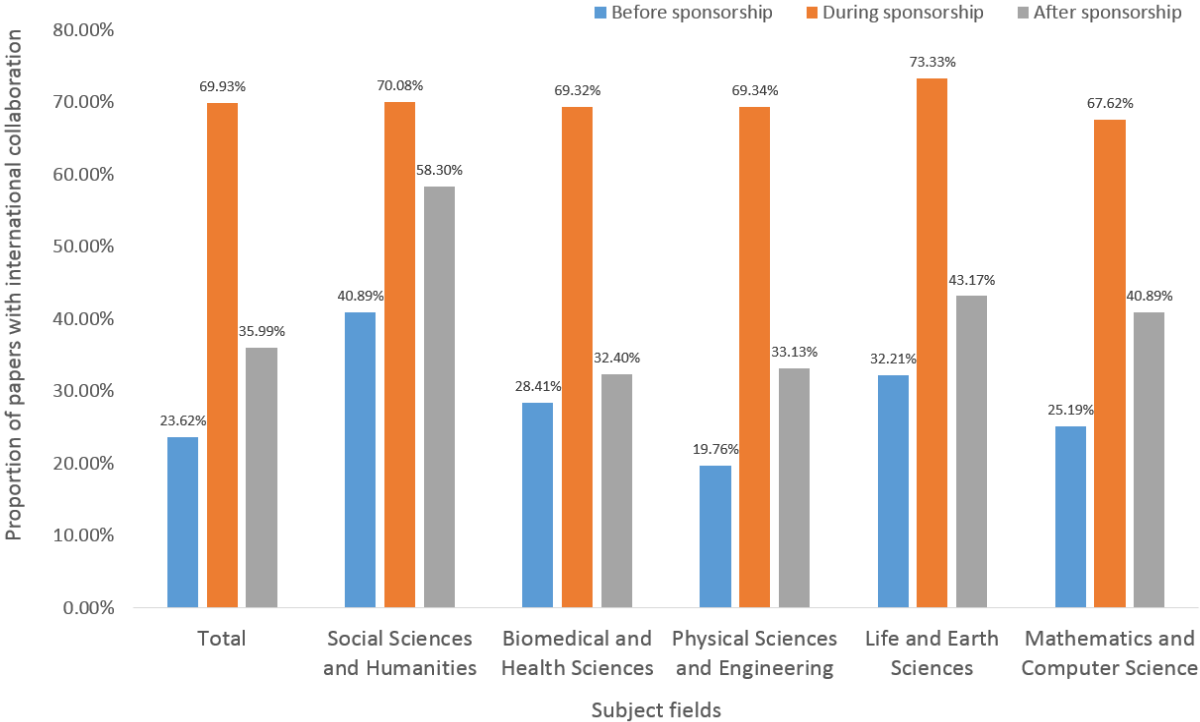

**Figure 4. Proportion of papers with international collaboration in different sponsorship periods.**

A similar pattern can be observed in total and across five main fields: papers published before sponsorship have the lowest rate of international collaboration, while papers published during sponsorship hold the highest proportion of international collaboration. After sponsorship, the proportion goes down compared to during sponsorship, but it is still higher than before sponsorship. These results suggest that CSC-funded scholars are most likely to collaborate with researchers from other countries during their mobility sponsored by the China Scholarship Council, and keep to some extent a relatively higher international collaboration after sponsorship. The retention rate of international collaboration after sponsorship varies across fields. For example, in the field of Social Sciences and Humanities, the proportion of papers with international collaboration during and after sponsorship are 70.1% and 58.3% respectively. In the field of Biomedical and Health Sciences, the proportion of papers with international collaboration after sponsorship decreased sharply, from 69.3% during sponsorship to 32.4%, although still being higher than the *before* period. The same downward trend can be found in other natural sciences and engineering sciences fields. One potential reason for the high proportion of international collaboration after sponsorship might be that most collaboration relationships established during mobility were continued, suggesting that the sponsorship from the China Scholarship Council contribute to improve the international collaboration from the perspective of research outputs, especially during the period when CSC-funded scholars conducted research abroad.

**Discussion and conclusions**

In this study, papers with funding from the China Scholarship Council were extracted from the Web of Science to investigate the field distribution of CSC-funded papers, the scientific mobility destinations of CSC-funded scholars, and the performance of the CSC scholarship in promoting international collaboration by providing awardees with the opportunity to conduct research abroad.

For the first research question about the productivity of papers contributed by CSC-funded scholars across fields, the results of the analysis presented in this paper shows that the number of papers funded by the China Scholarship Council increased over time. The CSC-funded papers are mainly from the fields of natural sciences and engineering sciences, especially physics and material science, chemistry and chemical engineering, which were emphasized by the China Scholarship Council in the outline for selecting awardees. However, social sciences and humanities, the key field that was highlighted by the China Scholarship Council in selecting scholarship awardees as well, has much fewer WoS-covered CSC-funded papers. There are several reasons for the lower production of CSC-funded papers in social sciences. In addition to the lower coverage and shorter acknowledgement index period in the Web of Science, another possible reason is that the number of CSC-funded scholars from the fields of natural sciences and engineering sciences is larger than that from the social sciences and humanities; thus the larger number of sponsored scholars from natural sciences would also explain the larger CSC-output in these fields.

Regarding the mobility destinations of CSC-funded scholars, through text-mining the acknowledgements of CSC-funded papers, a total of 9,467 scholars from mainland China sponsored by the CSC were identified. Thus, their mobility destinations can be tracked based on their affiliation information. On the basis of affiliation information of those identified Chinese CSC-funded scholars and their co-authors, most Chinese CSC-funded scholars chose to conduct research in the USA, Australia, Canada, and other developed European countries, such as Germany, the UK, and the Netherlands, and establish collaboration relationship with researchers from these countries.

Finally, for the third research question about the performance of the China Scholarship Council in promoting international collaboration, papers contributed by CSC-funded scholars during the sponsorship period show the highest proportion of international collaboration in whichever fields. During this period, scholars have moved abroad to conduct research, it is easier for them to interact and communicate with researchers in their mobility destinations and then expand their collaboration networks. Although the rate obviously declines for papers published after sponsorship, it is still higher than papers before sponsorship. Therefore, the support from the China Scholarship Council promoted the possibility for scholars to participate in international networks. This effect partly continued when the sponsorship had ended, since papers from CSC-funded scholars still showed a relatively higher proportion of international collaboration after the sponsorship.

There are several limitations to this study. Firstly, the same as previous studies focusing on scientific mobility by using bibliometric data, only the CSC-funded scholars with research outputs acknowledging the China Scholarship Council were identified and analysed, while those who never published WoS papers or didn't mention their funding sources are not considered, due to the lack of effective means of identifying these scholars and their performance based on bibliometric meta data. Secondly, for those papers not clearly mentioning the author who was sponsored by the China Scholarship Council in the acknowledgements, we cannot study which author is funded by the CSC scholarship. Thirdly, there might exist errors in the classification of sponsorship periods because of the publication delay. If a paper published before sponsorship was delayed for a long time, its publication date might be later than the last publication date of CSC-funded paper and the paper could be erroneously identified as *after sponsorship*. Lastly, as we mentioned above, the author name disambiguation algorithm has shortcomings in clustering Chinese names, although we have cleaned the authors' clusters data based on their Chinese first names and affiliations further, it is possible that there still exists inaccuracies.

**Acknowledgments**

This research is partially funded by the South African DST-NRF Centre of Excellence in Scientometrics and Science, Technology and Innovation Policy (SciSTIP). Zhichao Fang acknowledges the financial support from the China Scholarship Council (Grant No. 201706060201).

---

[1] Introduction to China Scholarship Council (in Chinese): https://www.csc.edu.cn/about

[2] Introduction to Chinese Government Scholarships: http://www.campuschina.org/content/details3_74776.html

[3] Provisions on the Management of Postgraduates Funded by China Scholarship for Studying Abroad (for Trial Implementation) (in Chinese): http://www.moe.edu.cn/s78/A20/s7068/201410/t20141021_178464.html

[4] Científicos lamentan estado de la ciencia en Perú (in Spanish): https://www.scidev.net/america-latina/migracion/noticias/cient-ficos-lamentan-estado-de-la-ciencia-en-per-.html

[5] Provisions on Several Issues Concerning the Dispatch and Management of Personnel Funded by China Scholarship for Studying Abroad (1996) (in Chinese): https://www.csc.edu.cn/chuguo/s/168

[6] 2018 Outline of selecting and dispatching CSC-funded scholars (in Chinese): https://www.csc.edu.cn/article/1042